\documentclass[%
reprint,
amsmath,amssymb,
prb,
]{revtex4-1}
\usepackage{graphicx}
\usepackage{dcolumn}
\usepackage{bm}
\usepackage[bookmarks]{hyperref}
\usepackage[version=3]{mhchem}
\usepackage[all]{hypcap}
\usepackage{float}
\usepackage{amsmath}
\usepackage{bm}
\usepackage[usenames,dvipsnames,svgnames,table]{xcolor}
\usepackage[center]{caption}

\begin{document}
	
	\preprint{APS/123-QED}
	
	\title{Effect of plasmonic Aluminum nanoparticles shapes on optical absorption enhancement in silicon thin-film solar cells}

	\author{Maedeh Rassekh$^{1}$}\email{rassekh@phd.guilan.ac.ir}
	\author{Reza Shirmohammadi$^{2}$}
	\author{Roghayeh Ghasempour$^{2}$}
	\author{Fatemeh Razi Astaraei $^{2}$}
	\author{Saber Farjami Shayesteh$^{1}$}
	
	\affiliation{$^{1}$Department of Physics, University of Guilan, 41335-1914 Rasht, Iran.\\
		$^{2}$Faculty of New Sciences and Technologies, University of Tehran, Iran.}
	
	\begin{abstract}
		Scattering from metal nanoparticles near their localized plasmon resonance; especially, the resonances of noble metals which are mostly in the visible or infrared part of the electromagnetic spectrum; is a way of improving light absorption in thin-film solar cells. The surface plasmon resonance can be affected by different factors such as the type, size, shape, and dielectric properties of the surrounding medium. Here we investigate, using the Finite Difference Time Domain (FDTD) method, how different shapes of aluminum nanoparticles affect absorption enhancement in silicon thin-film solar cells. Our results show that using these particles more than 30\% conversion efficiency for plasmonic solar cells can be achieved compared to a cell without particles. We have also found that although the spherical particles have the highest absorption peak, optimization of some parameters such as the height of the cylinder or disk-shaped particles and their distance from the substrate can increase the absorption. The results can provide more information and insight to understand and optimize plasmonic particles for solar cell applications.		
	\end{abstract}


	\maketitle
	
	
	\section{Introduction}	
	Over the past decades, using photovoltaic to produce electricity has been known as one of the promising ways of solving or at least decreasing the extreme speed of climate change which has become a defending issue in our time \cite{ranjbaran2019review}. In practice, however, the higher cost of this technology compared to fossil fuels has prevented their widespread use. Scientists always have been looking for a way to reduce the price and increase the efficiency of solar cells \cite{shiraz2018effect}. An ideal solar cell should have four main features: first, having a direct band-gap between 1.1 to 1.7 eV, second, consisting of non-toxic and available material in nature, third, having a high photovoltaic conversion efficiency, and four, to be produced easily for large areas \cite{chopra2004thin,goetzberger2002solar}.
			
	Among all the growing up kinds of solar cells \cite{steele2019thermal,wei2019lithium,ananthakumar2019third,darbe2019simulation,khatibi2019generation,yao2019plasmonic,khatibi2019optimization}, thin-film solar cells \cite{chopra2004thin} always have been considered as the low cost one. However, this cost reduction is due to the decrease in the thickness of the active layer, which is mostly made of silicon because of its cost-effective, earth abundance, non-toxicity, and processing availability \cite{li2010mode}. Reduction in the thickness of the active layer of solar cells can reduce efficiency by affecting their optical properties \cite{pillai2010plasmonics}. In general, the performance of a solar cell is determined by its ability to absorb light and generate electron-hole pairs. Both features can be limited by some factors such as light acceptance angle and spectral sensitivity of the active layer. The second one is responsible for the loss of a significant part of the solar spectrum \cite{shi2017enhancement}. To increase efficiency in this kind of solar cell, several techniques have been developed for better light absorption \cite{ghahremani2016high,liu2019effect,eisenhauer2019light,javadzadeh2019tunable,akdemir2020moox}. Plasmonic, as a new technique; that makes use of the nanoscale properties of metals; due to its potential to significantly enhance the cell's efficiency, has received great interest over the past decades \cite{kumawat2019indium,stuart1998island,rockstuhl2008absorption,beck2011light}. Enhancements in photocurrent have been observed for a wide range of semiconductors and solar cell configurations \cite{pillai2007surface,hagglund2008electromagnetic,derkacs2006improved}. To create a surface plasmon, two materials with free electrons at their interface are needed. In practice, this means that one of the two materials must be metal with free conduction electrons \cite{barnes2003surface}. The use of metal nanostructures such as gold and silver nanoparticles on the front surface, back surface, or inside the active layer can also excite the surface plasmon resonance \cite{hutter2004exploitation}. Incident electromagnetic radiation interaction with metallic nanoparticles results in collective electron oscillations, known as Localized Surface Plasmon Resonances (LSPRs). The result of this interaction is light field concentration and enhancement at visible and near-infrared wavelengths to the subwavelength scale \cite{rajeeva2015regioselective}. Surface plasmons excitations can increase light trapping and absorption and reduce reflection in thin substrates mainly by three methods: Scattering light into the device from the surface of metal nanoparticles situated on the front surface of the device that behave as bipolar (Far-field effect), near-field increment, and direct creation of charge carriers in the semiconductor layer \cite{atwater2011plasmonics}. The optical properties of metal nanoparticles are closely related to the real and imaginary parts of their dielectric constants, which depend on wavelength \cite{chan2008localized}. This means that the spectral response of the plasmonic nanoparticles is wavelength-dependent, and the optical properties of these particles can be tuned by changing some of their factors such as size, shape, and constituent material \cite{kelly2003optical}.
		
	It has been shown by S. H. Lim et al. that gold sphere nanoparticles with 100 nm in diameter positioned on top of a Si p-n junction photodiode can enhance light absorption as a result of the increment in electromagnetic field amplitude within the semiconductor \cite{lim2007photocurrent}. Finite-difference time-domain calculations for a single Ag or Au particle separated from a semi-infinite Si substrate by a 10 nm silicon dioxide layer by Catchpole et al. have shown that silver particles enhance the optical path length much higher than gold particles \cite{catchpole2008design}. T.L. Temple et al. have found that spherical silver nanoparticles with the same diameter suppress light absorption below the surface plasmon resonance frequency \cite{temple2009influence}. Akimov and Koh have demonstrated that aluminum nanoparticles result in higher optical absorption in the photoactive region than silver nanoparticles. They have found that Al is the best candidate among the commonly used metals due to its potential for non-resonant coupling and can double the enhancement compared to the resonant enhancement by small Ag nanoparticles for a wide range of surface coverage \cite{akimov2010resonant}. After that many experiments showed that Al nanoparticles are good candidates for frontside plasmonic scatterers since they give a broadband increase in the coupling of light into silicon \cite{villesen2012aluminum,uhrenfeldt2013tuning,uhrenfeldt2013diffractive,zhang2013improved,uhrenfeldt2015broadband}.
	
	The resonance of light inside the nanoparticles is also important because it can result in absorption of light by the particle, which will be lost as heat, and leads to a decrease in cell efficiency. In summary for nanoparticles positioned on the front surface, scattering the light mostly forward to the substrate is required. Absorbing by them and/or back-scattering will reduce efficiency. Thus, absorption enhancement in the substrate has to compensate these losses prior to cause a positive net gain. In the case of silver, its plasmonic resonance is in the solar spectrum. As a result, a significant absorption occurs by the nanoparticles themselves. In the case of aluminum, the resonance is outside the important part of the solar spectrum. In addition, aluminum particles are well oxidized, and their properties change slightly with shape and size. More importantly, their dispersion properties are stronger than that of silver nanoparticles \cite{akimov2011design}. However, it should be noted that aluminum and aluminum oxide particles exhibit toxic effects only in excessive concentrations, and hence, in general, their toxicity is considered low \cite{ambardar2020quantum,krewski2007human}.
	
		\begin{figure}[b]
			\includegraphics[width=1.0\columnwidth]{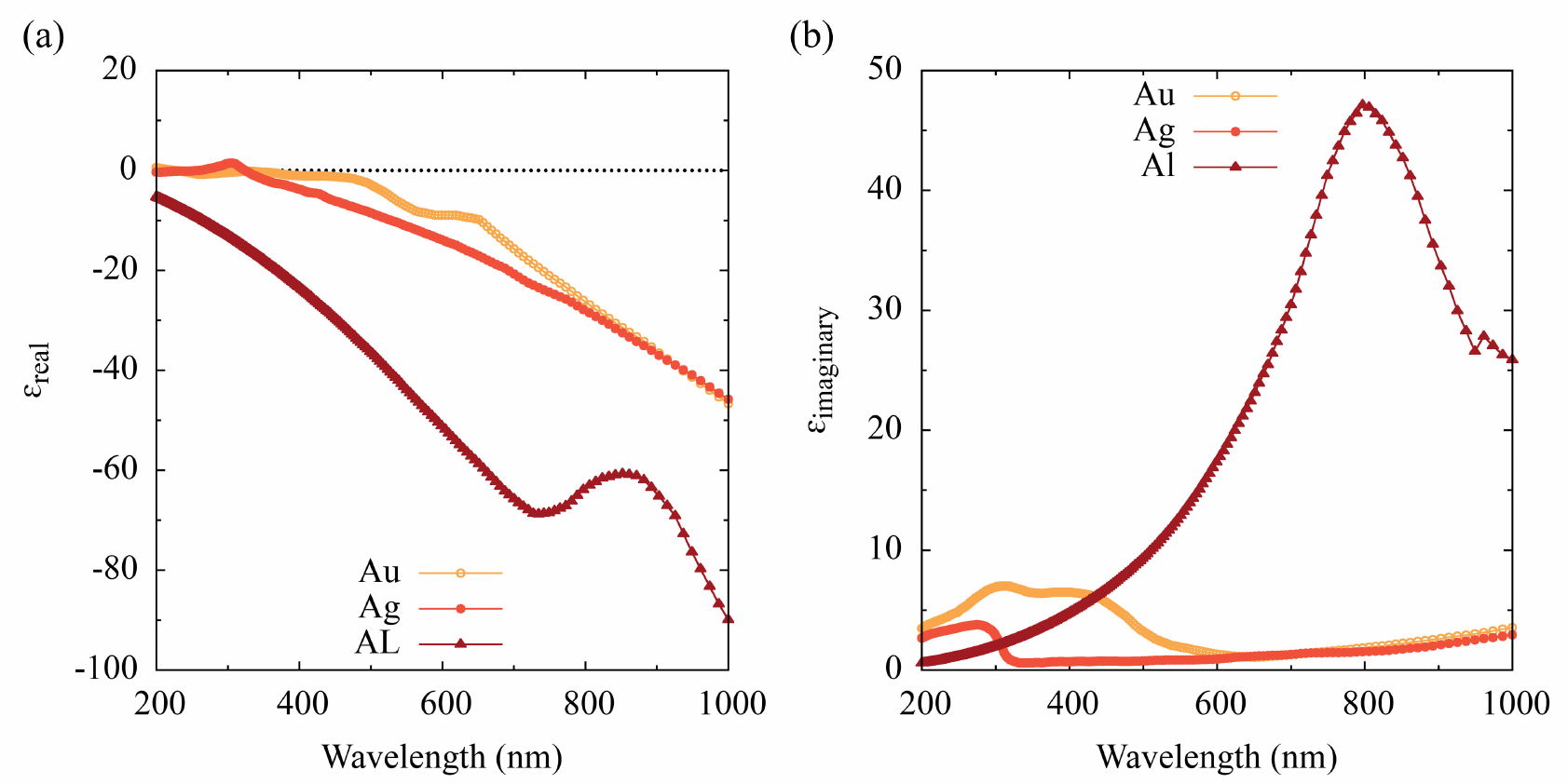}
			\caption{\small Comparison of the dielectric function of Au, Ag, and Al: real part (a), and imaginary part (b).}
			\label{fgr:1}
		\end{figure}	
	
	In general, to support plasmon excitation, the material needs to exhibit a large negative real and small positive imaginary dielectric function \cite{willets2007localized}. (Fig. \ref{fgr:1}) shows the real and imaginary parts of the dielectric functions of Au, Ag, and Al, as obtained from Palik \cite{edward1985handbook}. Au and Ag can only show plasmon excitations at wavelengths longer than about 450 nm and 350 nm, respectively. The real part of their dielectric function is positive below these wavelengths (Fig. \ref{fgr:1}-a). While Al should be plasmonically active from 200 nm to just below 800 nm. The Al peak at 800 nm in (Fig. \ref{fgr:1}-b) is due to the transition between two different bands associated with the conduction electrons (i.e., an interband transition) \cite{brust1970electronic}, which is accompanied by the rise in the imaginary part of the dielectric function. This makes Al less attractive at long wavelengths rather than Au and Ag. Nevertheless, Al has a remarkable potential over the visible and short wavelengths of the electromagnetic spectrum. In addition, the natural abundance of aluminum together with its low cost makes it attractive for practical applications \cite{kuiri2020control}.
			
	Recently, Mukti et al. have shown that spherical Al nanoparticles provide better light trapping compared with silver nanoparticles \cite{mukti2019increased}. In practice, we have nanoparticles in different sizes, shapes, and distances from the substrate. It has been shown by Cecilia Noguez \cite{noguez2007surface} that nanoparticles with fewer faces and sharper vertices exhibit resonances in a wider range of wavelengths. When a nanoparticle is truncated, the main resonance is blue-shifted, and the surface plasmon resonances at shorter wavelengths get closer to the dominant mode; so they can be hidden and increase the full width at half-maximum. Kelly et al. have shown that sharp edges increases the electric field enhancement. It has also been found that the corners contain more surface plasmons in a wider range of energy \cite{kelly2003optical}.
	
	Since the shape and size of a metallic nanoparticle dictate the spectral signature of its plasmon resonance, the ability to tune these parameters and study the effect on the plasmonic solar cells is an important challenge. Despite previous theoretical investigations on Al nanoparticles and their plasmonic properties, the literature on the effect of different parameters of these particles on the absorption in silicon-based thin-film solar cells is limited. So, in this work, we aim to reveal the effect of different parameters of aluminum particles such as their shape, height, and distance from the substrate on the absorption in silicon-based thin-film solar cells. Our paper is structured in the following manner. Details of the computational methodology are described in Section 2. We discuss our results in Section 3. Finally, conclusions and a summary of our results are given in Section 4.
	
	\section{Computational methods and theory}	
	
		\begin{figure}[t]
			\includegraphics[width=1.0\columnwidth]{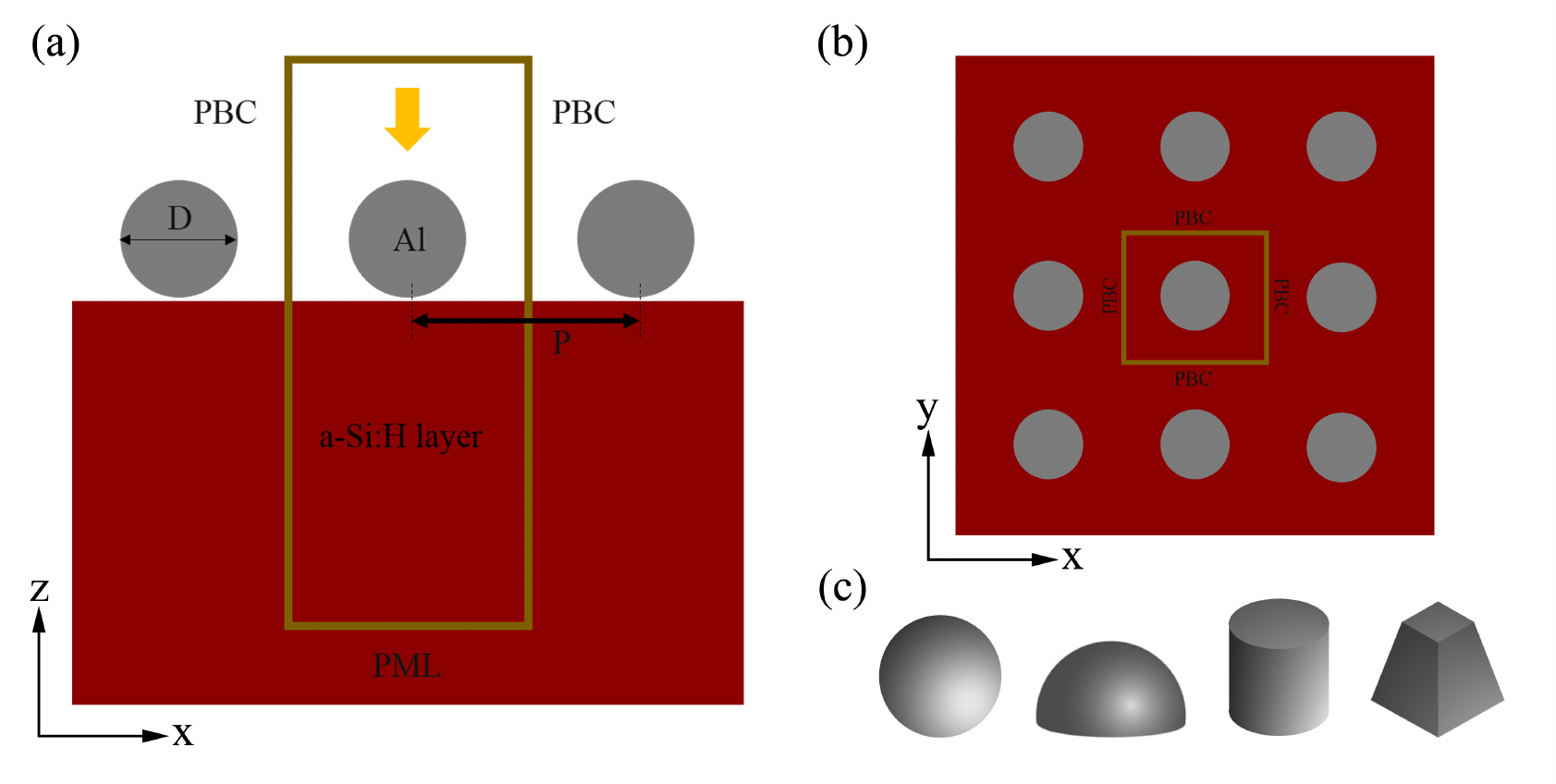}
			\caption{\small Schematic of the FDTD simulation model, consisting of a-Si:H layer, and aluminum (Al) nanoparticles thereon, exposed to the standard AM 1.5 solar spectrum. The Perfectly Matched Layer (PML) boundary conditions are used for upper and lower boundaries and Periodic Boundary Conditions (PBC) are used for the side boundaries to model the periodic nature of the particles. side view (a), top view (b), and considered shapes (c).}
			\label{fgr:2}
		\end{figure}
		\setlength{\textfloatsep}{10pt}	
							
	 We have performed the finite-difference time-domain (FDTD) calculations \cite{sullivan2013electromagnetic}. Which is a method for solving Maxwell’s equations in complex geometries, as implemented in the Lumerical FDTD Solutions \cite{FDTD}. The proposed structure, shown in (Fig. \ref{fgr:2}), consists of a crystalline silicon thin-film solar cell on which metal particles are distributed periodically. The particles are exposed to the standard AM 1.5 solar spectrum. To model the sunlight, a plane wave with a wavelength range from 400 nm to 1000 nm was normally incident (shown as the yellow arrow indicated in Fig. \ref{fgr:2}-a). The Perfectly Matched Layer (PML) boundary conditions are used for upper and lower boundaries, and Periodic Boundary Conditions (PBC) are used for the side boundaries to model the periodic nature of the particles. Symmetric and Anti-Symmetric boundary conditions have been used to reduce the required memory size and computation time. The thickness of Si is set to 800 nm, and the diameter and the period of particles are denoted as D and P, respectively. The refractive index of the surrounding medium is set to be 1. Experimentally, it is very likely that a native oxide layer is present above the Si surface and in contact with the Al nanoparticle. In the simulation, we have assumed that the Al particle are surrounded by vacuum, and their lowest point is 2 nm above the Si surface. It is well known that a thin oxide layer of about 2-3 nm thickness forms on the Al nanoparticles surface upon exposure to air \cite{pettit1975measurement,gertsman2005tem}. Since modeling a thin oxide layer in FDTD calculations requires sub-nanometer grid sizes and impractically long computing times, the natural oxide layer of Al is not taken into account. However, it has been shown by Akimov et al. that due to its transparency, $ Al_{2}O_{3} $ (whose dielectric permittivity is too low) cannot cause a significant change in the spectral absorption of Al nanoparticles \cite{akimov2011design}, whereas Ag nanoparticles exhibit unstable enhancement and due to their highly absorbing $ Ag_2O $ shell, they even result in a negative effect for certain ranges of oxide thicknesses \cite{akimov2010resonant}. The experimental study on the Al nanoparticles on a Si substrate without and with an oxide layer for silicon solar cell application performed by Piyush et al. also suggests that with the native oxide shell (2-3 nm) on the Al nanoparticles, the cell efficiency enhances compared to the cell without nanoparticles \cite{parashar2016plasmonic}.		
						
	The ratio of the number of carriers collected by a solar cell to the number of photons of a given wavelength incident on the solar cell is defined as the quantum efficiency of a solar cell $ QE(\lambda ) $ \cite{kalogirou2017mcevoy}. Using the quantum efficiency, integrated quantum efficiency , can be defined as \cite{li2010mode,singh2018optimized}:	
		\begin{equation}
			IQE=\frac{\int \frac{\lambda}{hc} QE(\lambda ) I_{AM1.5}(\lambda)d\lambda} {\int \frac{\lambda}{hc} I_{AM1.5}(\lambda)d\lambda}
		\end{equation}		
	where $ c $ is the speed of light in free space, $ h $ is Plank’s constant, and $ I_{AM1.5} $ is AM 1.5 solar spectrum. In this equation, the numerator and denominator mean the number of photons absorbed by the solar cell and that falling into the solar cell. To see how metal particles can improve the efficiency of the solar cell compared to a bare solar cell, we define the following quantities,
		\begin{equation}
			g(\lambda)=\frac{QE_{Particle}(\lambda)}{QE_{bare}(\lambda)}
		\end{equation}
	and,	
		\begin{equation}
		G=\frac{IQE_{Particle}}{IQE_{bare}}
		\end{equation}
	where $ g(\lambda) $ and $ G $ are enhancement in quantum efficiency and integrated quantum efficiency, respectively. Enhancement factors larger than 1 mean that the number of photons absorbed in the Si is increased with the nanoparticles compared to a bare Si surface.
	
	The absorption per unit volume is obtained from the divergence of the Poynting vector \cite{wu2014quantum},
		\begin{equation}\label{eq:6}
			P_{abs}=-\frac{1}{2} real(\vec{\triangledown}.\vec{P})
		\end{equation}
    the divergence of the Poynting vector can be defined as:
		\begin{equation}\label{eq:8}
		\vec{\triangledown}.\vec{P}=-\left( \vec{J}.\vec{E} + \frac{\partial u}{\partial t}\right)
		\end{equation}
	where J and u are charge density and energy density, respectively. Assuming that there is no free charge density, equation (\ref{eq:8}) can be written as:
		\begin{equation}
		\vec{\triangledown}.\vec{P}=- \frac{\partial u}{\partial t}
		\end{equation}
	and the energy density is given by:
		\begin{equation}
			u=\frac{1}{2} \left( \vec{E}.\vec{D}\right)=\frac{1}{2}\varepsilon|E|^2
		\end{equation}
	where D is the electric displacement field. Hence,
		\begin{equation}
		\vec{\triangledown}.\vec{P}=-\frac{1}{2} \frac{\partial}{\partial t} \left( \vec{E}.\vec{D} \right) 
		\end{equation}
	if: $ \vec{E}\propto e^{-i\omega t} $
		\begin{equation}\label{eq:12}
		\vec{\triangledown}.\vec{P}=-\frac{1 \times 2}{2} \varepsilon i \omega E^2 = i \omega \varepsilon E^2 
		\end{equation}		
	with the substitution of equation (\ref{eq:12}) into the equation (\ref{eq:6}), following equation yields \cite{singh2019plasmon}:
		\begin{equation}
		P_{abs}=-\frac{1}{2} real \left( i \omega \varepsilon E^2 \right) =  -\frac{1}{2} \omega |E|^2 imag (\varepsilon)
		\end{equation}
	Thus, by knowing the electric field intensity and the imaginary part of the permittivity, we can calculate the absorption as a function of space and frequency. Both of these quantities can be easily measured in an FDTD simulation.
	
		\begin{figure}[b]
			\includegraphics[width=1.0\columnwidth]{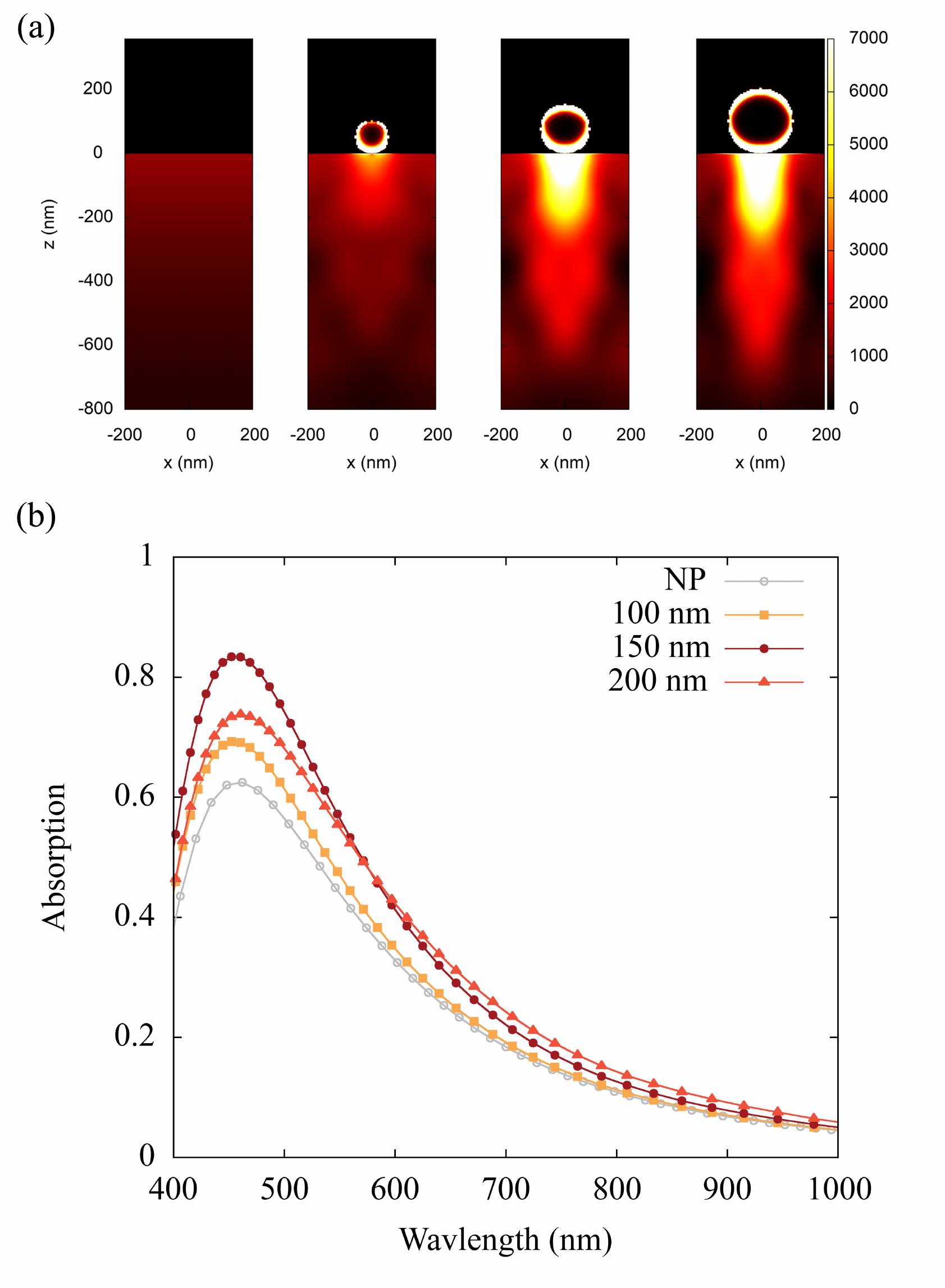}
			\caption{\small Absorption profile of x–z cross-section at wavelength of 500 nm in solar cells with aluminum particles with period 400nm and different diameters. From left to right: without particles, 50 nm, 100 nm, and 150 nm (a), and the Absorption spectra (b). “NP” indicates the absorption without particles.}
			\label{fgr:3}
		\end{figure}
			
	\section{Results and discussion}			

	We start with the spherical particles. (Fig. \ref{fgr:3}-a) shows the absorption profile of x–z cross-section in the silicon substrate and spherical Al particles with different diameters at a wavelength of 500 nm and a period of 400 nm. From left to right: without metal particles, particles with diameters of 100 nm, 150 nm, and 200 nm. The solar cell without nanoparticles is for comparative purposes. The larger the particle is, the stronger the absorption concentration and intensity becomes under the particle. The particle with a diameter of 200 nm shows the highest absorption under itself and at distances near to the surface of the substrate. Although the particle with a diameter of 150 nm shows less concentration of absorption under it, compared to a particle of 200 nm, they have increased the absorption in a wider part of the substrate. It can be concluded that the size of the deposited Al particles has a large influence on the absorption rate of the solar cell. A more detailed comparison is to see how the absorption increase at different wavelengths. (Fig. \ref{fgr:3}-b) shows the absorption characteristics of solar cells with the different diameters of particles in the 400 nm to 1000 nm wavelength range, demonstrating that the absorption of solar cells with Al particles increases significantly and that the peak value of absorption changes compared to that without Al particles. “NP” indicates the absorption without particles. For convenience, we use the normalized incident light power to unity. The distance between the particles (P) here is the same as 400 nm. It is clear that the aluminum particles increase absorption in the Si by the enhanced forward scattering due to the surface plasmon resonance. This enhancement is mainly in the wavelengths of 400 to 600 nm. Increasing the diameter from 100 to 150 nm shows an increment in the substrate absorption, but with a further increase from 150 to 200 nm, the absorption decreases. It is due to the more concentration of absorption under the particle and its less dispersion to other parts of the substrate. And confirmed that Al particles with a diameter of 150 nm provide the strongest absorption enhancement among the 3 different diameters of particles. This result indicates that the path length of incident light in the active layer can be enhanced by scattering short-wavelength electromagnetic waves by Al particles, thereby improving the optical absorption rate inside the photoactive layer. It should be noted that the calculation of the absorption spectra in FDTD is done by a more accurate numerical method compared to the method of calculating the absorption profile. And in these calculations, the error due to the discontinuity of the electric field near the interfaces is less.
		
		\begin{figure}[t]
			\includegraphics[width=1.0\columnwidth]{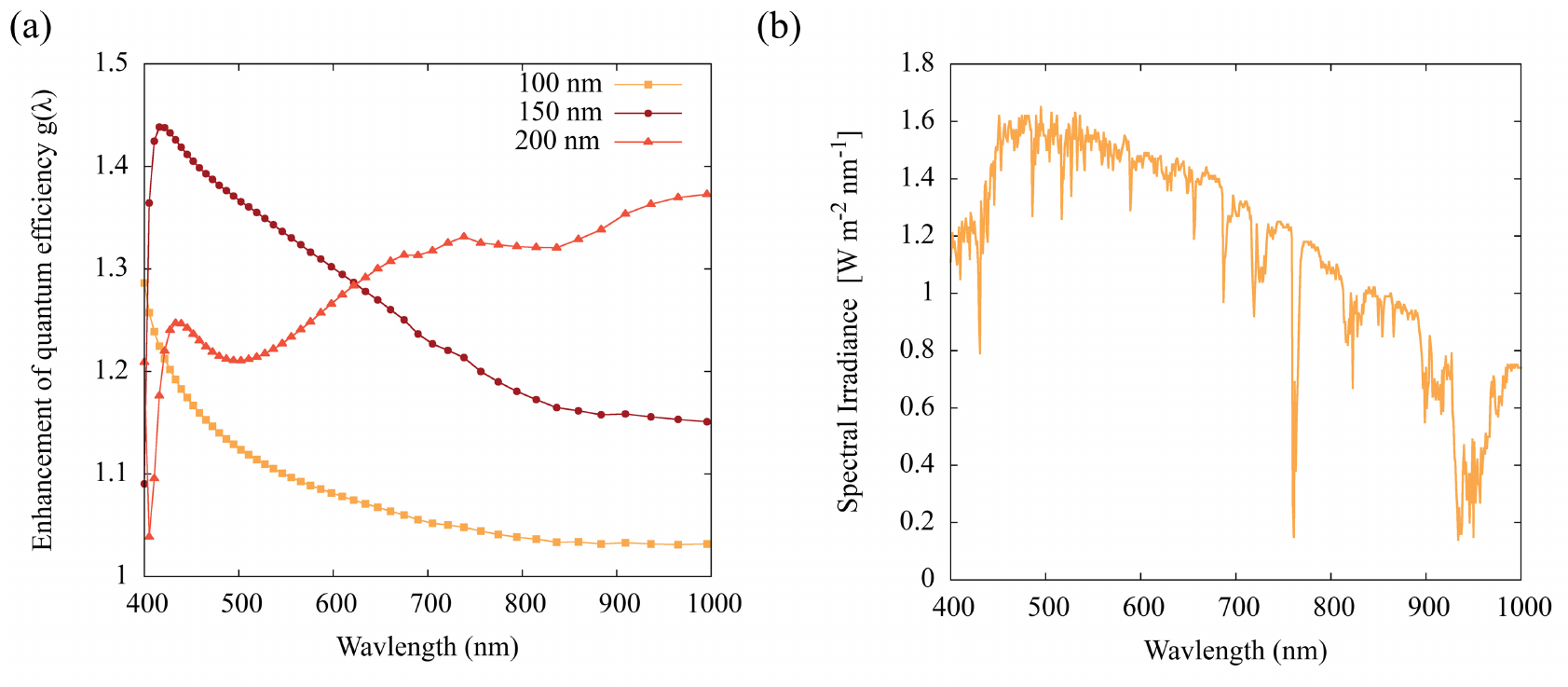}
			\caption{\small Enhancement of quantum efficiency $ g(\lambda) $ of solar cells with different diameters of Al spherical particles (a), and Sunlight spectrum (b).}
			\label{fgr:4}
		\end{figure}
		\begin{figure}[t]
			\includegraphics[width=1.0\columnwidth]{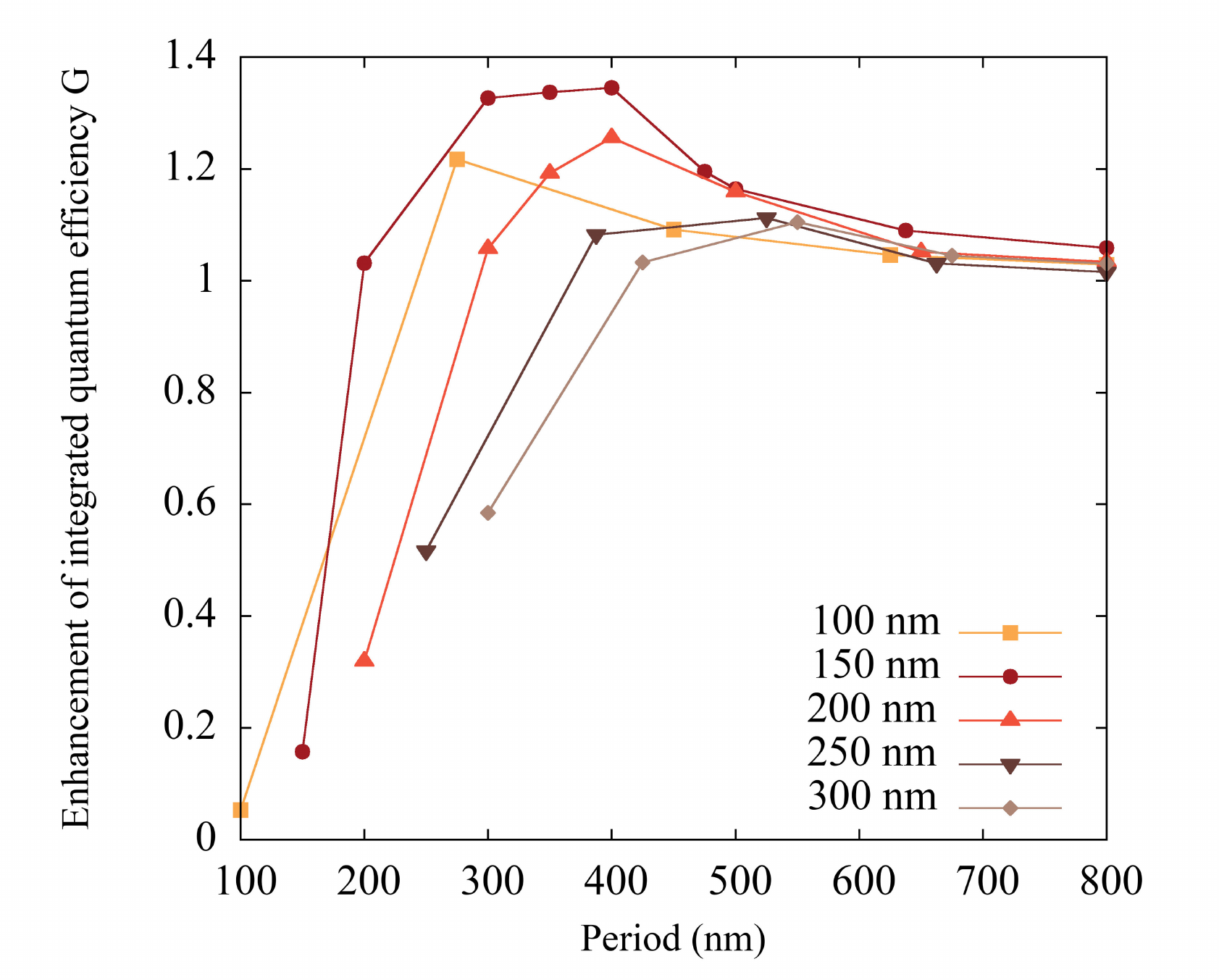}
			\caption{\small Enhancement of integrated quantum efficiency (G) of solar cells with different diameters and periods of Al spherical particles.}
			\label{fgr:5}
		\end{figure}
					
	To further investigate the effect of increasing the size of the particles, (Fig. \ref{fgr:4}) shows the increase in the quantum efficiency of the cell. All three particles show values larger than 1 in increasing the quantum efficiency of the cell, which means the particles increase the number of photons absorbed in the silicon layer. Particles with a diameter of 150 nm show the highest peak. And it shows a similar trend that we have already observed in increasing the absorption with increasing particle diameter in the previous plot. That is, by increasing the particle diameter from 100 to 150 nm, the peak of quantum efficiency increases, and then by increasing it further from 150 to 200 nm, the peak decrease remarkably. Particles with 150 nm in diameter show the most increment for wavelengths around 400 nm to 600 nm. As the particle size increase, the Al spectrum exhibits a redshift and broadening due to retardation effects and the emergence of higher-order plasmon modes. In addition, there is a trade-off between the absorptivity and the broadening of the absorption spectra. The broadening of the spectral line is at the expense of reducing the height of the peak.
	
		\begin{figure}[hb!]
			\includegraphics[width=1.0\columnwidth]{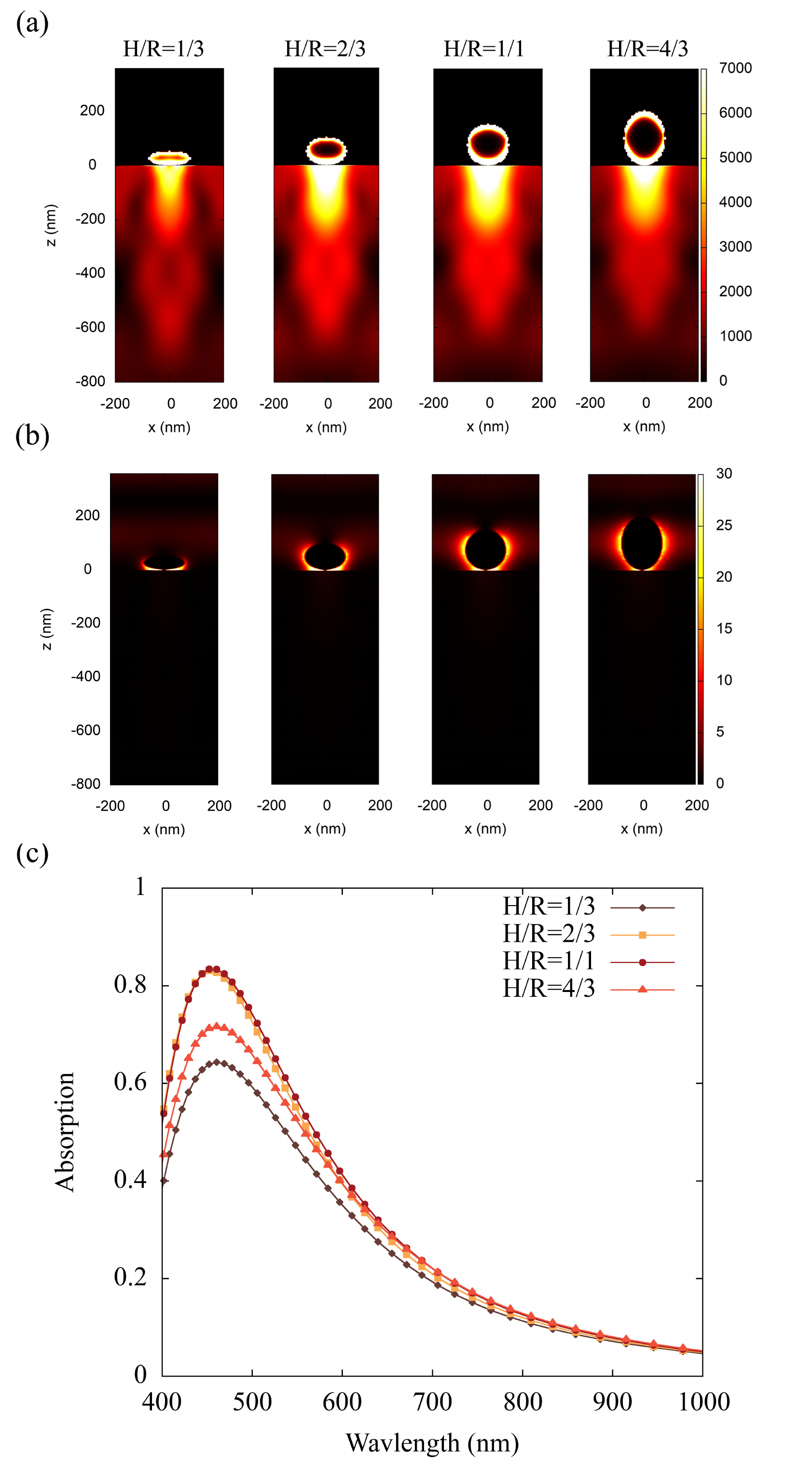}
			\caption{\small Absorption profile (a), electric field intensity profile (b) of x–z cross-section at wavelength of 500 nm for spherical particles with fixed radius R (75 nm) and different semi-heights H, and the Absorption spectra (c).}
			\label{fgr:6}
		\end{figure}
		
	In all calculations so far, the distance between the particles considered to be 400 nm. For more investigation, we calculate the increase in the total quantum efficiency of the solar cell as a function of the diameter and period of aluminum particles (Fig. \ref{fgr:5}). Enhancement of integrated quantum efficiency (G) takes minimum value when the period is equal to the diameter of the particle. It strongly increases with increasing the period until it reaches its maximum value. Adjacent particles exhibit strong coupling for periods slightly larger than the diameter of the particle \cite{rockstuhl2004analyzing}. Further Increase in the period causes a decoupling of the metallic structures, so the enhancement tends to 1 (no enhancement). This is because the surface of the Si is strongly diluted that means the number of particles per Si surface area decreases, so the particle enhancement effect can almost be disregard \cite{rockstuhl2008absorption}. Particles with a diameter of 150 nm show the largest increment in integrated quantum efficiency, and as the particle size increases further, the integrated quantum efficiency decreases. For these particles, periods between 300 and 400 nm have the largest increment in integrated quantum efficiency. As we can see in (Fig. \ref{fgr:5}), for particles with a diameter of 150 nm, there is a region where the conversion efficiency is substantially more than one. The reason is that the wavelength peak of g($ \lambda $) for these particles is close to the sun spectrum peak, which is around 500 nm (Fig. \ref{fgr:4}-b). From (Fig. \ref{fgr:5}), more than 30\% absorption-enhancement expected by using aluminum particles as a plasmonic solar cell compared to the bare solar cell.
	
		\begin{figure}[b]
			\includegraphics[width=1.0\columnwidth]{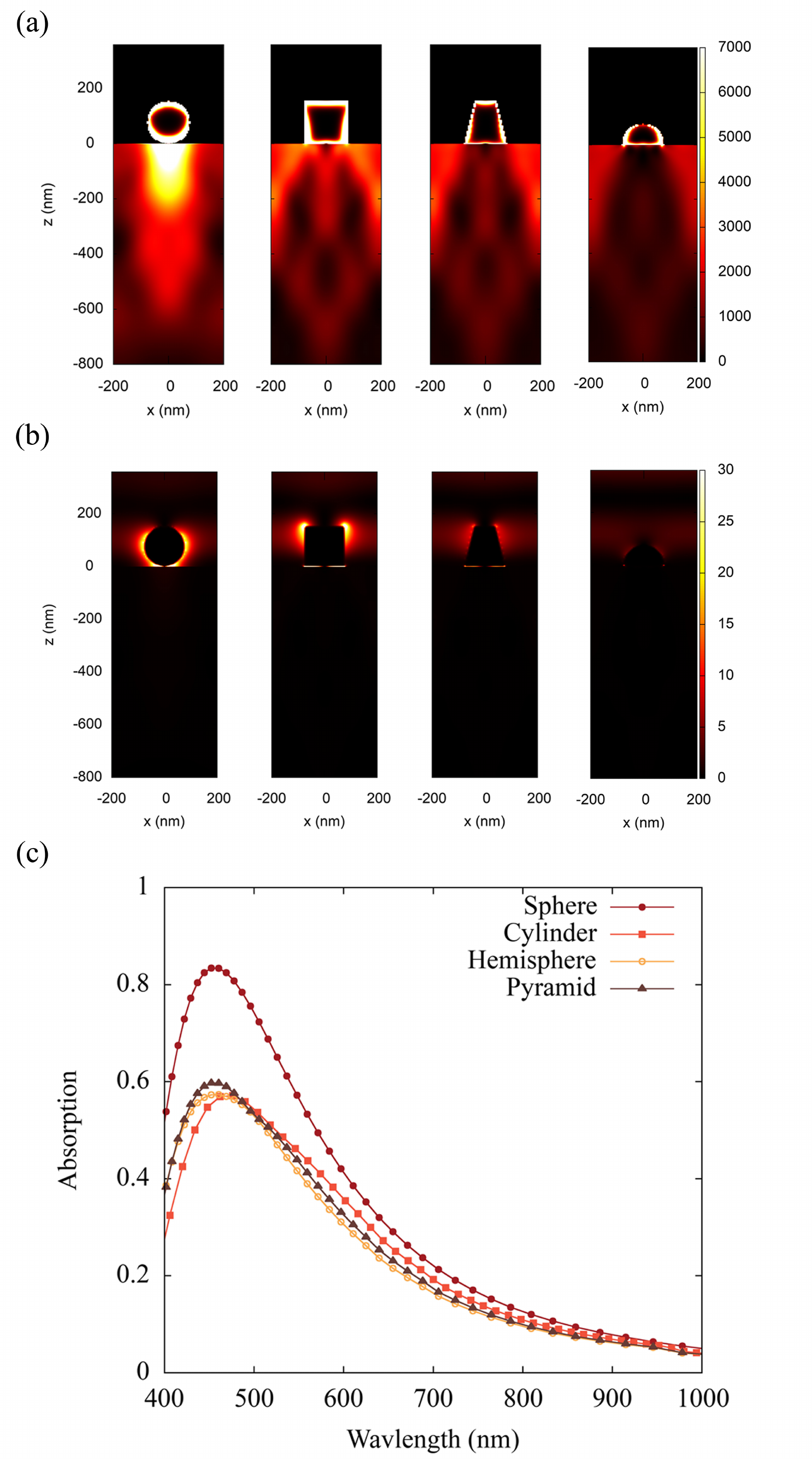}
			\caption{\small Absorption profile (a), electric field intensity profile (b) of x–z cross-section at wavelength of 500 nm for particles with different shapes. From left to right: Sphere, cylinder, hemisphere, and truncated pyramid, and the Absorption spectra (c).}
			\label{fgr:7}
		\end{figure}

	Here, we go further and study spherical particles with fixed radius R (75 nm) and different semi-heights H (Fig. \ref{fgr:6}) by varying the aspect ratio H/R in the range from 1/3 (oblate spheroids) to 4/3 (prolate spheroids). According to the obtained results, by increasing the ratio H/R, absorption first increases and then decreases, and it reaches its maximum when the ratio is equal to 1. For the oblate spheroidal shape with H/R = 2/3, the absorption peak is almost the same as particles with H/R = 1. Thus, for the particle size we have, oblate (H/R=2/3) and spherical shapes (H/R=1) are more preferable for the better performance of the cell. It worth mentioning that these results are in good agreement with previous results for Ag particles by Akimov et al. \cite{akimov2011design}.
				
	Considering the diameter of 150 nm and the period of 400 nm as the optimal diameter and period of the particles, we continue with other shapes. The distance of particles from the substrate is set to be 2 nm. For cylindrical or disk-shaped particles, the diameter and height of the particles are 150 nm. Particles with a truncated pyramid shape also have a height of 150 nm and a bottom and top x and y span of 150 nm and 75 nm, respectively. (Fig. \ref{fgr:7}-a) shows the absorption profile for aluminum particles with different shapes. As can be seen, spherical particles have the highest and hemispherical particles the least increase of absorption in the substrate. (Fig. \ref{fgr:7}-b) shows the intensity profile of the electric field for systems containing particles of different shapes. As expected, the scattering of the electric field also decreases similarly to the absorption. With the color bar limits that we have defined here for the electric field intensity, we can't see its dispersion inside the silicon layer. (Fig. \ref{fgr:7}-c) shows the absorption spectra for the solar cell in the presence of particles with different shapes. The highest absorption in the substrate belongs to the spherical particles. Other shapes of particles by shadowing have somewhat reduced the absorption in the substrate compared to the solar cell with no particles. However, since the particles in the form of cylinders or disks, after spherical particles, have shown the highest amount of absorption near the surface (Fig. \ref{fgr:7}-a), we continue with optimizing the effect of the height of these particles. But before that, we present a comparison for two other parameters for different shapes (Table. \ref{tbl:1}). The short circuit current density ($ J_{SC} $) is the electrical current generated by the solar cell when the voltage across the solar cell is zero. The $ J_{SC} $ can be calculated for both the bare Si substrate (no particles) and the Si substrate with the particles. The influence of different shapes of particles on the short circuit current was calculated for the entire wavelength from 400 nm to 1000 nm. For this analysis, the result shows a direct correlation to the absorption enhancement analysis shown previously, the spherical Al particles are giving the highest $ J_{SC} $ value. As we can see from (Table. \ref{tbl:1}), the $ J_{SC} $ increment for spherical particles compared to that of the bare silicon substrate is almost $ 34\% $. The other shapes show significantly lower $ J_{SC} $ values, with the hemisphere being the lowest, which its $ J_{SC} $ is even lower than that of the bare Si. The enhancement of integrated quantum efficiency (G) for the 400 nm period also follows a similar trend, with spherical Al particles producing the highest enhancement. This is in good agreement with the results of Choudhury et al. They had found that spherical Ag particles significantly increase the optical enhancement when compared to that of the other shapes \cite{choudhury2017influence}.
	
		\begin{table}[t!]
			\small
			\caption{\ The short circuit current density ($ J_{SC} $) and Enhancement of integrated quantum efficiency (G) for different shapes for the 400 nm period.}
			\label{tbl:1}
			\begin{tabular*}{0.48\textwidth}{@{\extracolsep{\fill}}ccc}
				\hline\\[-2ex]
				Shape & $ J_{SC} $ ($ mA/cm^{2} $) & G \\[0.1ex]
				\hline\\[-3pt]
				Bare Silicon & 5.77 & 1 \\
				Sphere & 7.73 & 1.34 \\
				Cylinder & 5.89 & 1.02 \\
				Truncated pyramid & 5.96 & 1.03 \\
				Hemisphere & 5.64 & 0.98\\			
				\hline
			\end{tabular*}
		\end{table}
	
		\begin{figure}[b]
			\includegraphics[width=1.0\columnwidth]{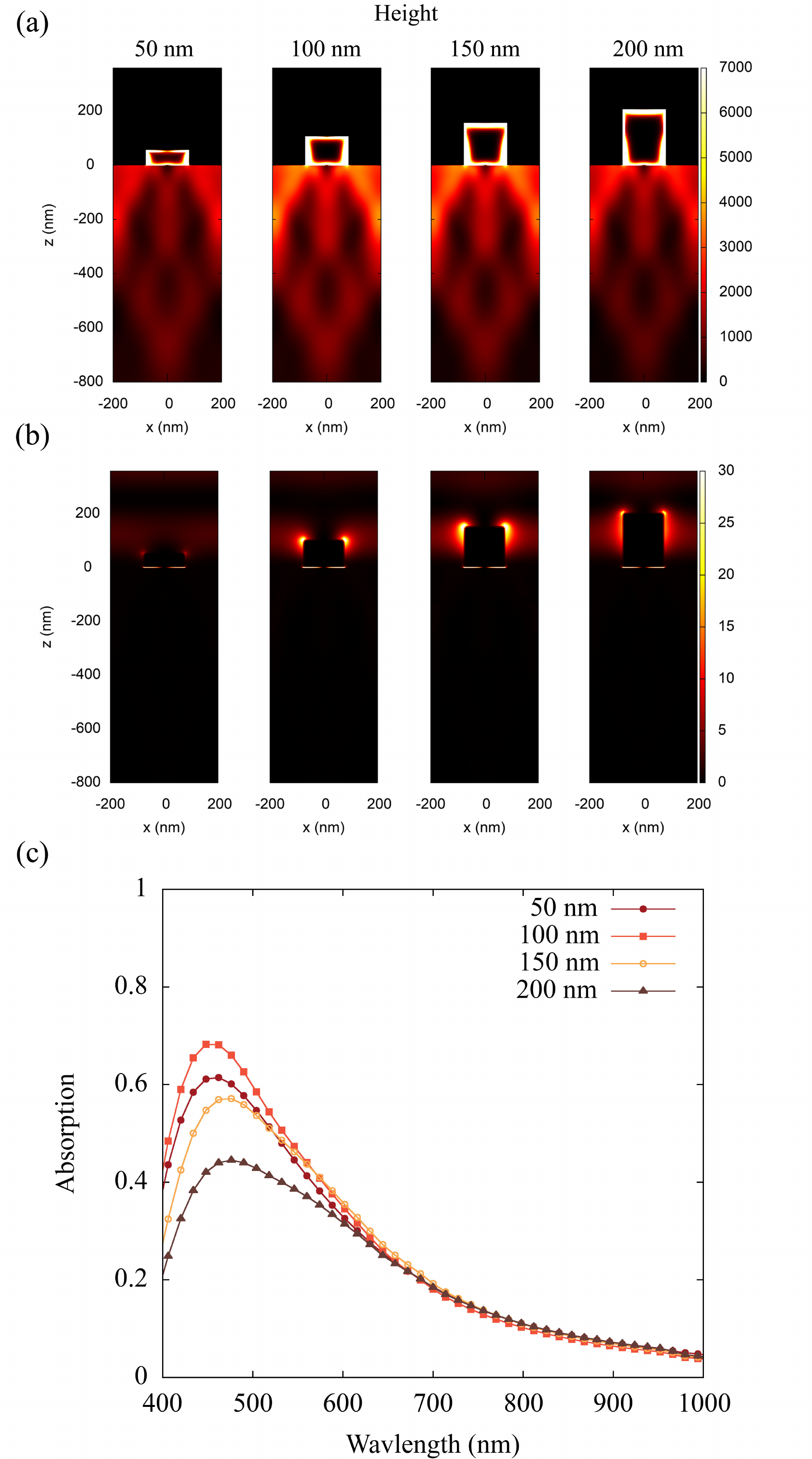}
			\caption{\small Absorption profile (a), electric field intensity profile (b) of x–z cross-section at wavelength of 500 nm for cylinder Al particles with different heights. From left to right: 50 nm, 100 nm, 150 nm, and 200 nm, and the Absorption spectra (c).}
			\label{fgr:8}
		\end{figure}
		\begin{figure}[t]
			\includegraphics[width=1.0\columnwidth]{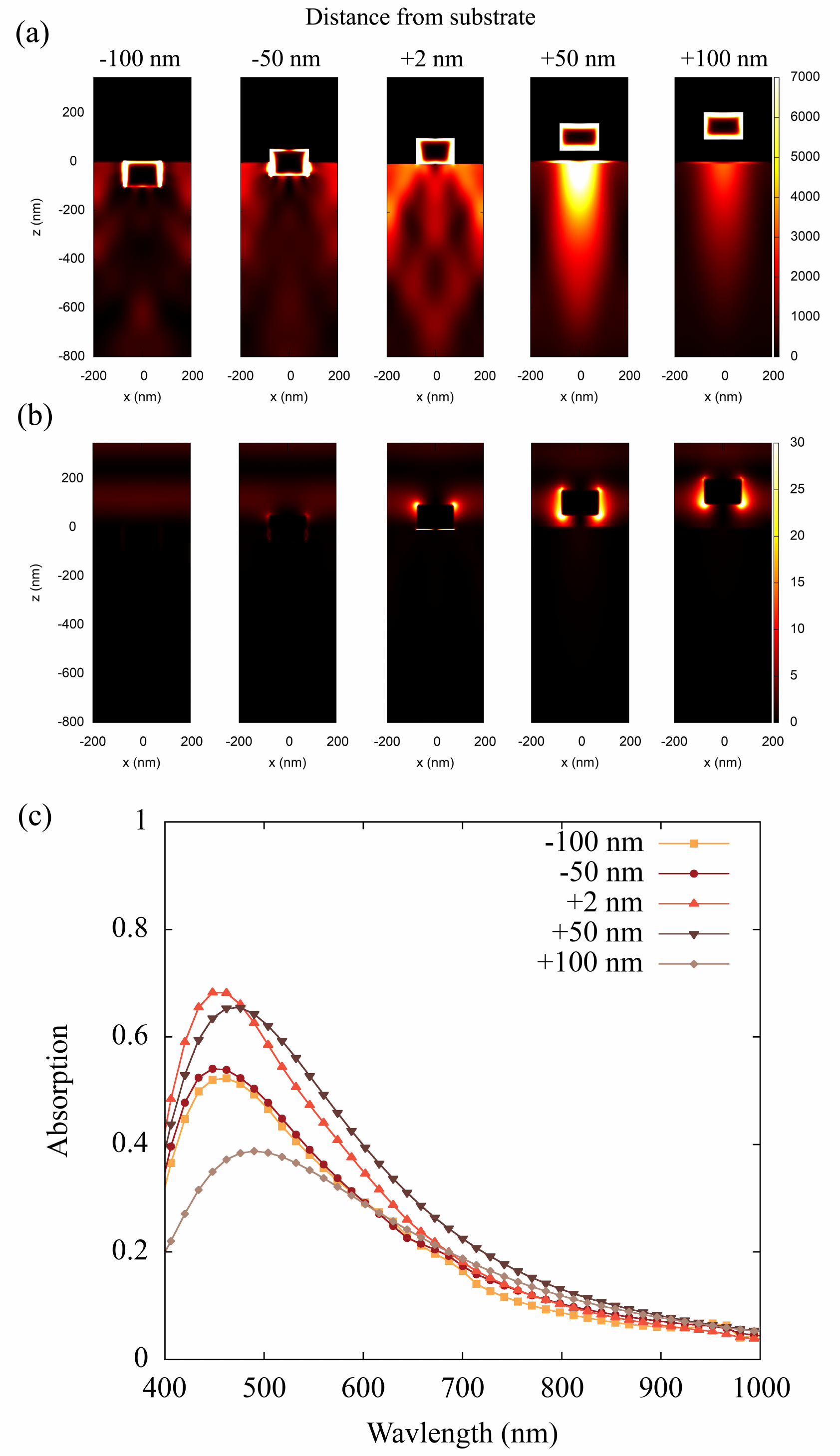}
			\caption{\small Absorption profile (a), electric field intensity profile (b) of x–z cross-section at wavelength of 500 nm for cylinder Al particles of 150 nm diameter and 100 nm height, with different distances from the substrate. From left to right: -100 nm, -50 nm, +2 nm, +50 nm, and +100, and the Absorption spectra (c). Negative values indicate the part of cylinders lie within the silicon substrate.}
			\label{fgr:9}
		\end{figure}
		
	(Fig. \ref{fgr:8}-a) shows the absorption profile for cylindrical or disk-shaped particles for different heights. As we can see in this figure, the absorption in the substrate increases with increasing particle height and then decreases with a further increase in height. The electric field profile (Fig. \ref{fgr:8}-b) shows that increasing the height causes the concentration of the electric field to decrease at the edges and to extend in the lateral circumference of the cylinder. From (Fig. \ref{fgr:8}-c), we conclude that although increasing the height from 50 nm to 100 nm, or in other words, changing the ratio of height to diameter from $ 1/3 $ to $ 2/3 $, increases the absorption in the substrate, but further increase in height reduces it. So that for the ratio of $ 4/3 $, the absorption has substantially decreased. Based on this, we choose the height of 100 nm for cylindrical particles with a diameter of 150 nm as the optimal calculated height, and in the following, we study the effect of the distance of the particles from the substrate.
	
		\begin{figure}[t]
			\includegraphics[width=1.0\columnwidth]{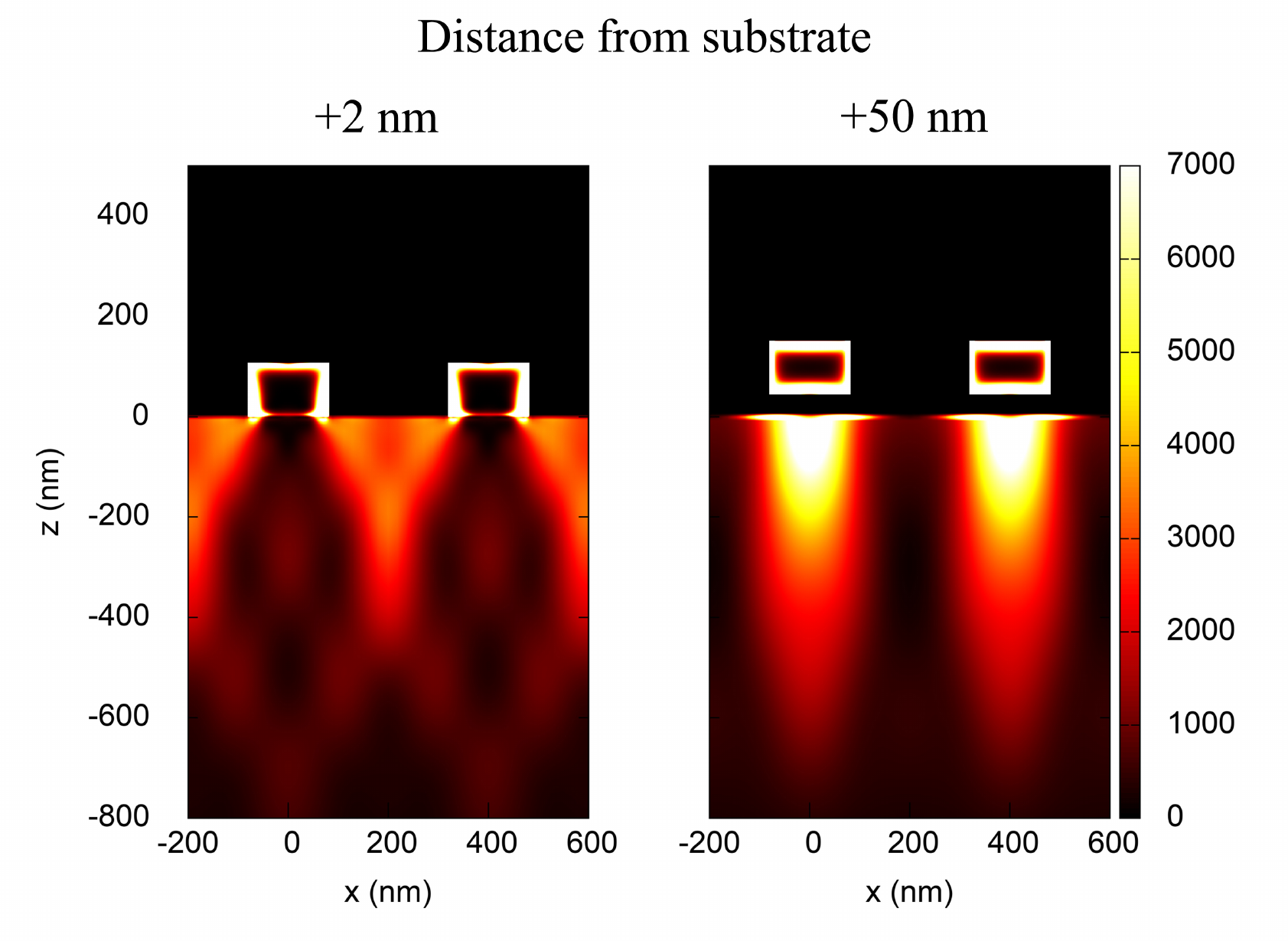}
			\caption{\small Absorption profile (a), electric field intensity profile (b) of x–z cross-section at wavelength of 500 nm for two cylinder Al particles of 150 nm diameter and 100 nm height with +2 nm and +50 nm distances from the substrate.}
			\label{fgr:10}
		\end{figure}
		
		\begin{figure}[b]
			\includegraphics[width=1.0\columnwidth]{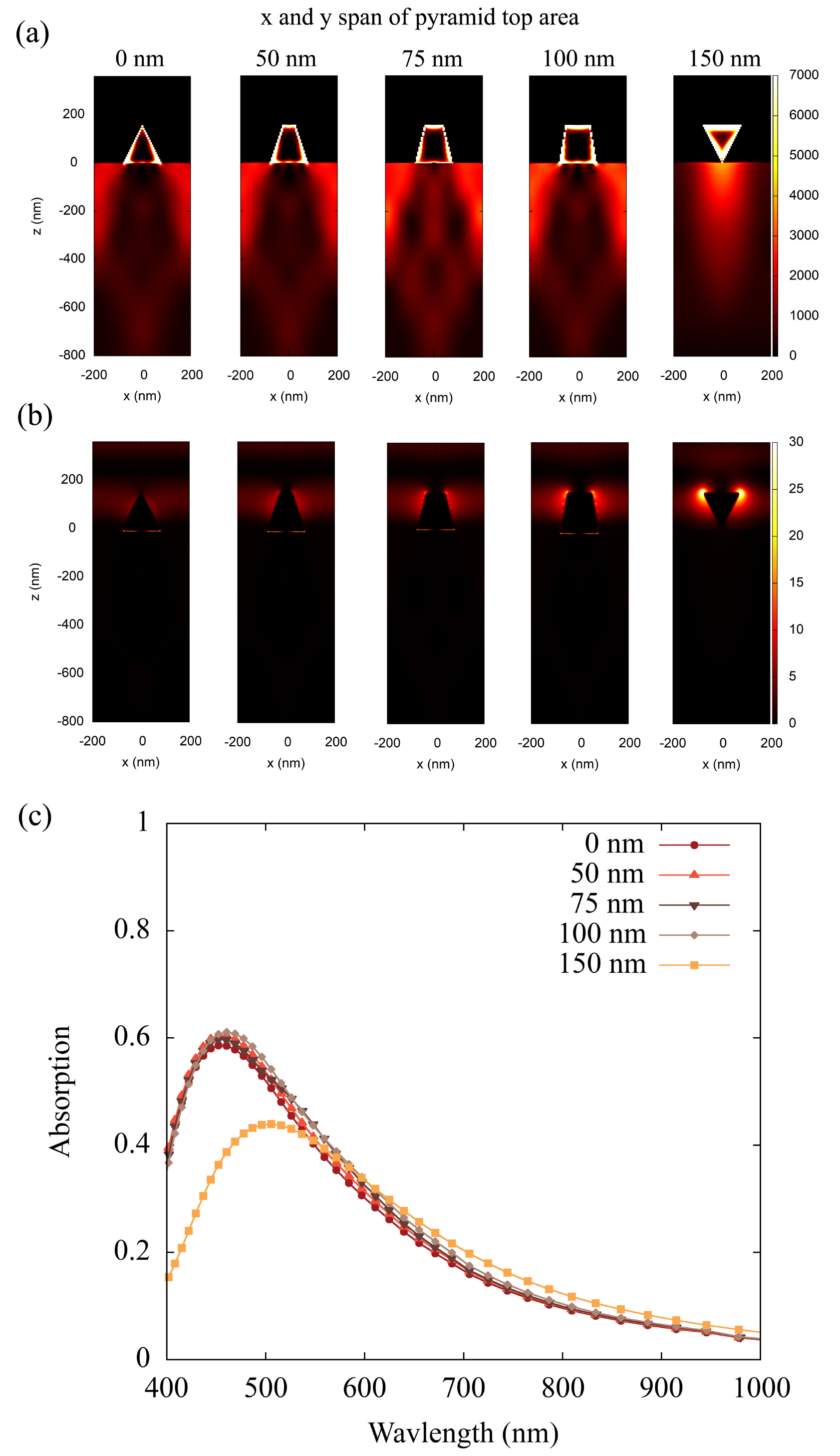}
			\caption{\small Absorption profile (a), electric field intensity profile (b) of x–z cross-section at the wavelength of 500 nm for pyramid Al particles of 150 nm height, and different x and y span of the top area. From left to right: 0 nm, 50 nm, 75 nm, 100 nm, and 150 nm (inverted pyramid), and the Absorption spectra (c).}
			\label{fgr:11}
		\end{figure}
		
	Absorption profiles for particles with different locations are shown in (Fig. \ref{fgr:9}-a). Five locations of Al particles were designed. From left to right include: placing the entire height of the particle inside the substrate, so that top of the particle is at the same level as the substrate surface, placing half of the height inside the substrate, placing the particles at a distance of 2 nm from the substrate (accordance to previous calculations), the particle at a distance of 50 nm, and at the distance of 100 nm from the substrate. For particles that are located at a distance from the substrate, a transparent layer that has a little absorption in the solar cell performance region (i.e., the visible wavelength) can be used as the separating layer. As we can see, when the whole particle height is placed inside the substrate, the absorption in the substrate is low, and by separating the particle from the substrate and placing it at 50 nm, the intensity and concentration of absorption under the particle inside the substrate, increases. As the particle gets further away from the substrate, absorption is still concentrated under the particle but decreases in intensity. The cross-section of the electric field intensity (Fig. \ref{fgr:9}-b) shows that as the distance from the substrate increases, the concentration of the electric field changes from the edges to the lateral circumference. And further increment shifts it to the lower edges of the particles. Similar results have been observed by F. J. Beck et al. for Ag nanodisks \cite{beck2011resonant}. For a more detailed study, we plot the absorption spectra in terms of wavelength for different distances of particles from the substrate. (Fig. \ref{fgr:9}-c) shows that the absorption for the particles with a 2 nm distance from the substrate is almost equal to the absorption for the particles with a 50 nm distance from the substrate. This is because although the particles at a distance of 50 nm from the substrate show a more increment in absorption, compared to the particles that are located at a distance of 2 nm from the substrate, most of this increase has concentrated under the particle and the coupling between the adjacent particles has almost gone (Fig. \ref{fgr:10}). Therefore, this makes the area where the absorption increases in total to be a smaller part of the substrate. Increment in absorption under the particles comes at the cost of reducing the area in which absorption has increased. So, in (Fig. \ref{fgr:9}-c), we see that the absorption peaks of these two are close to each other.
		
	For the last part, we investigate pyramid Al particles with different x and y span of its top area. As we can see in (Fig. \ref{fgr:11}), changing the x and y span of the top area does not have much effect on absorption. However, as it expands and changes to a more cubic shape, it shows a very little increase. Inverting the pyramid causes the absorption to be more concentrated under the particle and the coupling between the particles loss. In addition, it increases the absorption of the particle itself.
	
	\section{Conclusions}
	In this paper, we demonstrated that particle shape is a crucial parameter for designing plasmon enhanced solar cells. We investigated, through the time-domain finite-difference (FDTD) method, a thin-film crystalline silicon solar cell on which Al particles have distributed periodically. Various parameters, including absorption profile and the electric field intensity profile of x–z cross-section for different shapes, were studied. The enhancement of integrated quantum efficiency was calculated as a function of particle diameter and period. The structures were compared with bare thin-film Si solar cells. Simulation results suggest that compared to a bare cell, more than $ 30\% $ conversion efficiency for plasmonic solar cells can be achieved using Al spherical particles. Spherical particles with a diameter of 150 nm and a period of 400 nm show the greatest effect on the absorption. While in most of the previous works, particles were treated as a single particle, so the coupling between them was ignored, we considered the coupling between the particles. So the situation can be more close to the experiments.

\end{document}